\documentclass[reprint,amsmath,amssymb,aps]{revtex4-1}
\usepackage{graphicx} 
\usepackage{datetime}
\usepackage{dcolumn} 
\usepackage{bm} 
\usepackage{xcolor}
\DeclareMathOperator\arctanh{arctanh}
\DeclareMathOperator\arccosh{arccosh}

\begin{document}
\title{Landau damping of surface  plasmons in metal nanostructures}
\author{Tigran V. Shahbazyan}
\affiliation{
Department of Physics, Jackson State University, Jackson, MS 39217 USA
}


\begin{abstract} 
We develop a quantum-mechanical theory for Landau damping  of surface plasmons in metal  nanostructures  of arbitrary shape. We show that the  electron surface scattering, which facilitates plasmon decay in small nanostructures, can be incorporated into the metal dielectric function \textit{on par} with  phonon and impurity scattering. The derived surface scattering rate is determined by the  local field  polarization relative to the metal-dielectric interface and is highly sensitive to the system geometry. We illustrate our model by providing analytical results for surface scattering rate in  some common shape nanostructures. Our results can be used for calculations of hot carrier generation rates in photovoltaics and photochemistry applications.
\end{abstract}
\maketitle

\section{Introduction}

Surface plasmons are collective electron excitations that provide unprecedented means for energy concentration, conversion, and transfer at the nanoscale \cite{atwater-jap05,ozbay-science06,stockman-review}. Plasmons can be resonantly excited in metal-dielectric nanostructures  giving rise to strong oscillating local fields that underpin numerous plasmon-enhanced spectroscopy phenomena, including surface-enhanced Raman scattering \cite{sers}, plasmon-enhanced fluorescence  and energy transfer \cite{novotny-book}, or plasmonic laser (spaser) \cite{bergman-prl03}. Among   key characteristics that  impact many plasmonics applications \cite{lakowicz-ab01,duyne-nm06,atwater-nm10,nordlander-nn15} is the plasmon lifetime, which, depending on the plasmonic system size, is governed by several decay mechanisms \cite{halperin-rmp86,kresin-pr92,vallee-jpcb01,schatz-jpcb03,noguez-jpcc07}.  While in large systems,  the plasmon lifetime is mostly limited by radiation \cite{liao-prl82}, in  systems with characteristic size $L<c/\omega$, where $c$ and $\omega$ are, respectively, the light speed and frequency, the dominant decay mechanism is excitation of electron-hole (\textit{e-h}) pairs by the plasmon local field  accompanied by  phonon and impurity scattering or, for small systems, surface scattering  \cite{kreibig-book}. Recently, plasmon decay into \textit{e-h} pairs has attracted intense interest as a highly efficient way of hot carrier generation and  transfer across the interfaces with   applications in photovoltaics \cite{park-nl11,melosh-nl11,halas-science11,halas-nc13,lian-nl13,halas-nl13-2,clavero-np14,brongersma-nl14,atwater-nc14,halas-nc15} and photochemistry \cite{brongersma-nl11,moskovits-nl12,halas-nl13,moskovits-nn13,halas-jacs14}.   Plasmon-assisted hot carrier generation is especially efficient in smaller plasmonic systems, where light scattering is relatively weak and  extinction is dominated by resonant plasmon absorption. In such systems, carrier excitation   is enhanced due to  strong surface scattering that provides  a new momentum relaxation channel \cite{kreibig-book}.

Surface-assisted plasmon decay (Landau damping) has been extensively studied  experimentally \cite{klar-prl98,mulvaney-prl02,halas-prb02,klar-nl04,vallee-prl04,halas-nl04,hartland-pccp06,vallee-nl09,vanduyne-jpcc12,vallee-nl13,schatz-nl15} and theoretically \cite{kawabata-jpsj66,lushnikov-zp74,schatz-jcp83,barma-pcm89,yannouleas-ap92,eto-srl96,uskov-plasmonics13,khurgin-oe15,jalabert-prb02,jalabert-prb05,yuan-ss08,vallee-jpcl10,lerme-jpcc11,li-njp13,nordlander-acsnano14,mortensen-16} since the pioneering paper by Kawabata and Kubo \cite{kawabata-jpsj66}, who have shown that, for a spherical  particle  of radius $a$, the surface scattering rate is $\gamma_{sp}=3v_{F}/4a$, where $v_{F}$ is the electron Fermi velocity. In subsequent quantum-mechanical studies carried within random phase approximation (RPA) \cite{lushnikov-zp74,schatz-jcp83,barma-pcm89,yannouleas-ap92,eto-srl96,uskov-plasmonics13,khurgin-oe15} and time-dependent local density approximation (TDLDA) \cite{jalabert-prb02,jalabert-prb05,yuan-ss08,vallee-jpcl10,lerme-jpcc11,li-njp13,nordlander-acsnano14,mortensen-16} approaches, a more complicated picture has emerged involving the role of confining potential and nonlocal effects. These are dominant at the spatial scale $\xi_{nl}=v_{F}/\omega$ that defines the characteristic length for nonlocal effects \cite{mortensen-pn13,mortensen-nc14} (e.g., for noble metals, $v_{F}/\omega <1$ nm  in the plasmon frequency range), whereas for larger systems with $L\gg v_{F}/\omega$ (i.e., several nm and larger), they mainly affect the overall magnitude of $\gamma_{sp}$, while preserving intact its size dependence \cite{lerme-jpcc11,nordlander-acsnano14}. The latter implies that in a wide size range $v_{F}/\omega \ll L\ll c/\omega$, which includes most  plasmonic systems used in applications, the detailed structure of electronic states is unimportant, and the confinement effects can be reasonably described in terms of electron surface scattering, which can be incorporated, along with phonon and impurity scattering, in the metal dielectric function $\varepsilon(\omega)=\varepsilon'(\omega)+i\varepsilon''(\omega)$. Here, we adopt the Drude  dielectric function $\varepsilon(\omega)  = \varepsilon_{i}(\omega)-\omega_{p}^{2}/\omega(\omega+i\gamma)$, where $\varepsilon_{i}(\omega)$ describes interband transitions, $\omega_{p}$ is the plasma frequency, and $\gamma$ is the scattering rate. Thus, it is expected that, for  systems in the above size range, the scattering rate should be modified as $\gamma=\gamma_{0}+\gamma_{s}$, where $\gamma_{0}$ is the bulk scattering rate and $\gamma_{s}$ is the \textit{surface scattering rate}. In particular, the standard expression, in terms of metal dielectric function, for the plasmon decay rate \cite{stockman-review},
\begin{equation}
\label{ld}
\Gamma = 2\varepsilon''(\omega)\left [\frac{\partial \varepsilon'(\omega)}{\partial \omega}\right ]^{-1},
\end{equation}
should   describe plasmon damping due to \textit{both} bulk and surface-assisted processes if surface-modified $\varepsilon(\omega)$ is used instead. For example, for $\omega$  well below the onset of interband transitions, the rate (\ref{ld})  coincides with (modified) Drude scattering rate: $\Gamma \approx \gamma=\gamma_{0}+\gamma_{s}$.

The major roadblock in the way of carrying this program forward has so far been the lack of any  quantum-mechanical model for evaluation of $\gamma_{s}$ in a nanostructure of \textit{arbitrary} shape. Due to the complexity of electronic states in general-shape confined systems, calculations of $\gamma_{s}$ were performed, within RPA \cite{kawabata-jpsj66,lushnikov-zp74,schatz-jcp83,barma-pcm89,yannouleas-ap92,eto-srl96,uskov-plasmonics13,khurgin-oe15} and TDLDA \cite{jalabert-prb02,jalabert-prb05,yuan-ss08,vallee-jpcl10,lerme-jpcc11,li-njp13,nordlander-acsnano14,mortensen-16} approaches, only for some simple (mostly spherical) geometries. For general shape systems, the surface scattering rate was suggested, within the \textit{classical scattering} (CS) model \cite{kreibig-zp75,ruppin-pss76,schatz-cpl83,schatz-jcp03,moroz-jpcc08}, in the form $\gamma_{cs}=A  v_{F}/L$, where $L$ is interpreted as the ballistic scattering length in a classical cavity, while the phenomenological constant $A$ accounts for the effects of surface potential, electron spillover, and dielectric environment. However, the unreasonably wide range  of measured $A$ ($0.3\div 1.5$  for  spherical particles \cite{kreibig-book}) raised questions about the CS model validity \cite{dionne-nature12}, while  recent measurements of plasmon spectra in  nanoshells \cite{halas-nl04}, nanoprisms \cite{vanduyne-jpcc12}, nanorods \cite{vallee-nl13}, and nanodisks \cite{schatz-nl15}  revealed  significant discrepancies with its predictions. Furthermore, the CS approach is questionable on physical grounds as well since it involves carrier scattering  across the entire system even for $L\gg v_{F}/\omega$, i.e.,  when the nonlocal effects are expected to  be weak.

On the other hand, surface scattering should depend sensitively on the local fields accelerating the carriers towards  the metal-dielectric interface. This dependence was, in fact, masked in all  previous quantum-mechanical studies of simple-shape systems \cite{kawabata-jpsj66,lushnikov-zp74,schatz-jcp83,barma-pcm89,yannouleas-ap92,eto-srl96,uskov-plasmonics13,khurgin-oe15,jalabert-prb02,jalabert-prb05,yuan-ss08,vallee-jpcl10,lerme-jpcc11,li-njp13,nordlander-acsnano14,mortensen-16},  where a specific functional form of the local field, appropriate for the given geometry, was adopted, while it is completely missing in the CS approach. Moreover, for the most widely studied spherical geometry, the local field is uniform inside the particle (apart from surface effects), which further obscured its importance. Note, however, that our  recent RPA calculations of the surface plasmon lifetime in spherical metal nanoshells with dielectric core \cite{kirakosyan-prb16} revealed the crucial role of local fields; for thin shells, the field is pushed out of the metal region, resulting in  a reduction of the plasmon decay rate. This result contrasts sharply with the CS model predictions but, in fact, is consistent with the measured light-scattering spectra of single nanoshells \cite{halas-nl04}. Furthermore, recent measurements of plasmon spectra in nanorods and nanodisks revealed strong sensitivity of plasmon modes' linewidth to the local field polarization relative to the system symmetry axis \cite{schatz-nl15}. For general-shape systems, the local field orientation relative to the interface can strongly affect the surface scattering rate, and, therefore, must be properly accounted for in any consistent theory of surface-assisted plasmon decay. 

\begin{figure}[tb]
\begin{center}
\includegraphics[width=0.9\columnwidth]{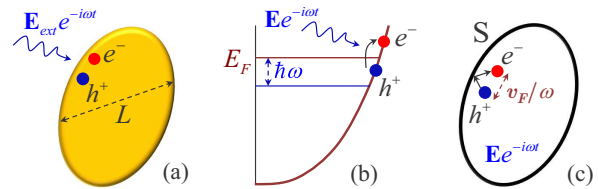}
\caption{\label{fig1}
Schematics for surface-assisted excitation of an \textit{e-h} pair with energy $\hbar\omega$. (a) An external optical field incident on a metal nanostructure of characteristic size $L$, (b) excites a surface plasmon that decays into an \textit{e-h} pair, (c) accompanied by momentum relaxation via carrier surface scattering in a small region of size $v_{F}/\omega\ll L$.
  }
\end{center}
\vspace{-6mm}
\end{figure}

In this paper, we present a quantum-mechanical theory for surface-assisted \textit{e-h} pair excitation by   alternating local electric field $\textbf{E}e^{-i\omega t}$ in metal nanostructures of general shape. 
We note that intraband absorption of energy $\hbar\omega$ takes place in a region of  size $v_{F}/\omega$ (see Fig.~\ref{fig1}) and, therefore, can be viewed as a \textit{local} process in systems with characteristic  size $L\gg v_{F}/\omega$. We show that, within RPA, surface scattering can be included into the Drude dielectric function by modifying the scattering  rate as  $\gamma=\gamma_{0}+\gamma_{s}$.  We derive the surface scattering rate $\gamma_{s}$ as
\begin{equation}
\label{rate-surface}
\gamma_{s}=
A\,v_{F} \, \frac{\int \! dS |E_{n}|^{2}}{\int \! dV |\textbf{E}|^{2}},
\end{equation}
where $E_{n}$ is the local field component \textit{normal} to the interface and the integrals  are carried  over the metal surface (numerator) and volume (denominator). The constant $A$ has the value $A=3/4$  for hard-wall confining potential, but can be adjusted to account for surface and nonlocal  effects. The full plasmon decay rate, including  bulk \textit{and} surface contributions, has still  the form (\ref{ld}), but with modified $\varepsilon(\omega)$ that now includes the surface scattering rate  (\ref{rate-surface}). Surface scattering is highly sensitive to the local field polarization relative to the metal-dielectric interface, leading to distinct rates for different plasmon modes, as we illustrate here for some common  geometries. 

%
%

The paper is organized as follows. In Sec.~\ref{sec:general}, we outline our   approach to   plasmon Landau damping in metal nanostructures. In Sec.~\ref{sec:power}, we derive an explicit expression for surface-assisted absorbed power and the corresponding scattering rate in systems of arbitrary shape. In Sec.~\ref{sec:rates}, we present analytical and numerical results for surface scattering rates in some common nanostructures. In Sec.~\ref{sec:conc}, we discuss the effect of confining potential profile on the surface scattering rate, and  the Appendices detail some technical aspects of our calculations.

%
\section{Decay rate of surface plasmons in metal nanostructures}
\label{sec:general}

In this section, we outline our approach to calculation of  the plasmon  decay rate for a metal nanostructure embedded in dielectric medium.  For simplicity, we restrict ourselves by  metal structures occupying some volume $V$ with a single surface $S$, so that the local dielectric function $\varepsilon (\omega,\bm{r})=\varepsilon' (\omega,\bm{r})+i\varepsilon'' (\omega,\bm{r})$  equals $\varepsilon(\omega)$ and $\varepsilon_{d}$  in the metal and dielectric regions, respectively. For systems with characteristic size $L\ll c/\omega$, the retardation effects are unimportant, and plasmon modes are determined by the Gauss law $\nabla \cdot \left [\varepsilon' (\omega_{l},{\bm r}) {\bf E}_{l}({\bm r})\right ]=0$,
where ${\bf E}_{l}({\bm r})$ is the  slow component of  plasmon local field  and $\omega_{l}$ is the plasmon mode frequency. For brevity, we omit the mode index $l$ hereafter. The general expression for plasmon decay rate $\Gamma$ has the form \cite{shahbazyan-prl16}
\begin{equation}
\label{ld-formal}
\Gamma =  \frac{Q}{U},
\end{equation}
where $U$ is the mode energy \cite{landau},
\begin{align}
\label{energy-LL}
U=\frac{\omega}{16\pi} \frac{\partial \varepsilon'(\omega)}{\partial \omega}\! \int\! dV|\textbf{E}|^{2},
\end{align}
and $Q$ is the absorbed power (loss function)
\begin{equation}
\label{power}
Q=\frac{\omega}{2}\,\text{Im}\! \int \! dV \textbf{E}^{*}\cdot \textbf{P}.
\end{equation}
Here, $\textbf{P}({\bm r})$ is the electric polarization vector and the star stands for complex conjugation. In the classical (local) picture, the polarization vector is proportional to the local field, $\textbf{P}_{loc}(\bm{r})=\textbf{E}(\bm{r})[\varepsilon(\omega,\bm{r})-1]/4\pi$, yielding  the absorbed power due to the bulk processes \cite{landau}
\begin{equation}
\label{power-bulk}
Q=\frac{\omega\varepsilon''(\omega)}{8\pi} \! \int \! dV  |\textbf{E}|^{2},
\end{equation}
which, along with the mode energy (\ref{energy-LL}), leads to the standard form (\ref{ld}) of the plasmon  damping rate.

Surface contribution to the absorbed power, $Q_{s}$, comes from  the momentum relaxation channel provided by carrier scattering from the metal-dielectric interface. Since   surface scattering introduces nonlocality, $Q_{s}$ must be evaluated microscopically. The general expression for  $Q_{s}$ can be obtained by relating $\textbf{P}(\bm{r})$ to the electron polarization operator $P(\omega;\bm{r},\bm{r}')$  via the induced charge density:
\begin{equation}
\label{induced}
\rho(\bm{r})=e\! \int \!  d{\bm r}'P(\bm{r}, \bm{r}') \Phi(\bm{r}')=-\bm{\nabla}\cdot {\bf P}(\bm{r}),
\end{equation}
%
where local potential $\Phi (\bm{r})$ is  defined as $e{\bf E} (\bm{r})=-\bm{\nabla} \Phi (\bm{r})$ ($e$ is the electron charge).  With the help of Eq.~(\ref{induced}), integration of Eq.~(\ref{power}) by parts yields
\begin{equation}
\label{power1}
Q_{s}=\frac{\omega}{2}\,\text{Im} \! \int \! dV dV' \Phi^{*}(\bm{r})P(\omega;\bm{r},\bm{r}') \Phi(\bm{r}'),
\end{equation}
where $P(\omega;\bm{r},\bm{r}')$ includes only the electronic contribution, i.e., without phonon and impurity scattering effects. Within RPA,  $P(\omega;\bm{r},\bm{r}')$ is replaced by the polarization operator for noninteracting electrons \cite{mahan-book}, yielding
\begin{equation}
\label{power-rpa}
Q_{s}=\pi \omega\sum_{\alpha\beta}|M_{\alpha\beta}|^{2}
\left [f(\epsilon_{\alpha})-f(\epsilon_{\beta})\right ]
\delta(\epsilon_{\alpha}-\epsilon_{\beta}+\hbar\omega),
\end{equation}
where $M_{\alpha\beta}=\int dV \psi_{\alpha}^{*}\Phi\psi_{\beta}$ is the transition matrix element of local potential $\Phi(\bm{r})$ calculated from the wave functions  $\psi_{\alpha}(\bm{r})$ and $\psi_{\beta}(\bm{r})$ of electron states with energies $\epsilon_{\alpha}$ and $\epsilon_{\beta}$ separated by $\hbar\omega$,  $f(\epsilon)$ is the Fermi distribution function, and   spin degeneracy is accounted for.

In terms of $Q_{s}$, the surface-assisted contribution to the plasmon decay rate, i.e., the Landau damping (LD) rate, has the form 
\begin{equation}
\label{rate-surface-gen}
\Gamma_{s}=\frac{Q_{s}}{U},
\end{equation}
where $U$ is given by Eq.~(\ref{energy-LL}). Note that often in the literature, $\Gamma_{s}$ is  identified with the standard first-order transition probability rate, given by the expression similar to Eq.~(\ref{power-rpa}) but divided by the factor  $\hbar\omega/2$. We stress that in a system with dispersive dielectric function, where the mode energy is $U$ rather than $\hbar\omega$ \cite{landau}, the standard transition rate must by rescaled by  the factor $\hbar\omega/2U$ \cite{shahbazyan-prl16}.

Calculation of $Q_{s}$ (and, hence, of $\Gamma_{s}$) hinges on the transition matrix element $M_{\alpha\beta}$, which has so far been evaluated, either analytically or numerically, only for several simple geometries permitting  separation of variables \cite{kawabata-jpsj66,lushnikov-zp74,schatz-jcp83,barma-pcm89,yannouleas-ap92,eto-srl96,uskov-plasmonics13,khurgin-oe15,jalabert-prb02,jalabert-prb05,yuan-ss08,vallee-jpcl10,lerme-jpcc11,li-njp13,nordlander-acsnano14,kirakosyan-prb16}. For  general-shape systems,  evaluation of $M_{\alpha\beta}$ presents an insurmountable challenge of finding, with a good accuracy, the three-dimensional electron wave functions oscillating rapidly,  with the Fermi wavelength period $\lambda_{F}$, on the system size scale $L\gg \lambda_{F}$. However, as we demonstrate in the following section, this difficulty can be bypassed and even turned into an advantage as $Q_{s}$ is derived in a closed form for any nanostructure larger than the nonlocality scale, i.e.,  for $L\gg v_{F}/\omega$. 



\section{Absorbed power and surface scattering rate}
\label{sec:power}

In this section, we derive the surface contribution $Q_{s}$ to the absorbed power due to \textit{e-h} pair excitation by alternating local field $\textbf{E}e^{-i\omega t}$ created in the metal either by plasmons or as a response to an external   field.

We start with the transition matrix element $M_{\alpha\beta} = \int\! dV \psi^{\ast}_{\alpha}\Phi \psi_{\beta}$, where $\psi_{\alpha}(\bm{r})$ is the eigenfunction of the Hamiltonian $H=-(\hbar^{2}/2m)\Delta$ for an electron with energy $\epsilon_{\alpha}$ in a hard-wall cavity (this approximation is discussed later).  We  consider the case when excitation energy $\hbar\omega$  is much larger than the electron level spacing, so that, in the absence of phonon and impurity scattering, the electron transition to the state $\psi_{\beta}(\bm{r})$  with energy $\epsilon_{\beta}=\epsilon_{\alpha}+\hbar\omega$  requires momentum transfer to the interface. A direct evaluation of  $M_{\alpha\beta}$,  so far carried out only for some simple geometries \cite{kawabata-jpsj66,lushnikov-zp74,schatz-jcp83,barma-pcm89,yannouleas-ap92,eto-srl96,uskov-plasmonics13,khurgin-oe15,kirakosyan-prb16,jalabert-prb02,jalabert-prb05,yuan-ss08,vallee-jpcl10,lerme-jpcc11,li-njp13,nordlander-acsnano14}, requires the knowledge of $\psi_{\alpha}$ in the entire system volume. We note, however, that for a typical plasmon frequency $\hbar\omega\ll E_{F}$, where $E_{F}$ is the Fermi energy in the metal, the momentum transfer $q\sim \hbar\omega /v_{F}$ takes place in a region of size $\xi_{nl}\sim \hbar/q\sim v_{F}/\omega$, so that, for characteristic system size $L\gg v_{F}/\omega$, the \textit{e-h} pair excitation takes place in a close proximity to the interface (see Fig.~\ref{fig1}).  It is our observation that, for an electron in a hard-wall cavity, the boundary contribution to $M_{\alpha\beta}$ can be  extracted as a surface integral of the form,
\begin{equation}
\label{matrix}
M_{\alpha\beta}^{s}= \frac{-e\hbar^{4}}{2m^{2} \epsilon_{\alpha\beta}^{2}} \! \int \! dS [\nabla_{n}\psi_{\alpha}(\bm{s})]^{*}E_{n}(\bm{s}) \nabla_{n}\psi_{\beta}(\bm{s}),
\end{equation}  
where $\nabla_{n}\psi_{ \alpha}(\bm{s})$ is the wave-function normal derivative at a surface point $\bm{s}$, $E_{n}(\bm{s})$  is the corresponding normal field  component, $\epsilon_{\alpha\beta}=\epsilon_{\alpha}-\epsilon_{\beta}$ is the \textit{e-h} pair excitation energy, and $m$ is the electron mass. The derivation of Eq.~(\ref{matrix}) is given in Appendix \ref{app:a}. Using the above matrix element, Eq.~(\ref{power-rpa}) can be  recast as
\begin{equation}
\label{Q-surface}
Q_{s}=\frac{e^{2}\hbar^{4}}{4\pi m^{4} \omega^{3}} \! \int \! \! \int \!  dS dS' E_{n}(\bm{s})E_{n'}^{\ast}(\bm{s}') F_{\omega}(\bm{s},\bm{s}'),
\end{equation}
where $F_{\omega}(\bm{s},\bm{s}')$ stands for \textit{e-h} surface correlation function defined as
\begin{equation}
\label{F}
F_{\omega}(\bm{s},\bm{s}')=\! \int \! d\epsilon   f_{\omega}(\epsilon) \rho_{nn'}(\epsilon;\bm{s},\bm{s}')\rho_{n'n}(\epsilon+\hbar\omega;\bm{s}',\bm{s}).
\end{equation}
Here, the function $f_{\omega}(\epsilon)=  f(\epsilon)-f(\epsilon+\hbar\omega)$ restricts the electron initial energy  to the interval $\hbar\omega$ below $E_{F}$, and 
\begin{equation}
\rho_{nn'}(\epsilon;\bm{s},\bm{s}')=\nabla_{n}\nabla'_{n'}\text{Im} G(\epsilon;\bm{s},\bm{s}')
\end{equation}
is the normal derivative of  the electron cross density of states $\rho(\epsilon;\bm{r},\bm{r}')=\text{Im} G(\epsilon;\bm{r},\bm{r}')$ at surface points, where
\begin{equation}
G(\epsilon;\bm{r},\bm{r}')=\sum_{\alpha} \frac{\psi_{\alpha}(\bm{r})\psi_{\alpha}^{*}(\bm{r}')}{\epsilon -  \epsilon_{\alpha}+i0}
 \end{equation}
is the  Green function of a confined electron. Note that neither the Green function $G(\epsilon;\bm{r},\bm{r}')$ nor the correlation function $F_{\omega}(\bm{s},\bm{s}')$  can be evaluated  with any reasonable accuracy for a general-shape cavity. However, an explicit expression for $Q_{s}$ in terms of local fields  can still be derived by exploiting the  difference in the length scales characterizing  electron and plasmon excitations. Namely, while the electron wave-functions oscillate with the Fermi wave length period $\lambda_{F}$, the local fields  significantly change  on the much larger system scale $L\gg \lambda_{F}$. Below we outline the main steps of our derivation of $Q_{s}$ and refer   to Appendix \ref{app:b} for details.
 

First, we note that since excitation of an \textit{e-h} pair with energy $\hbar\omega$ near the Fermi level takes place in  a region of  size $v_{F}/\omega$,  the correlation function $F_{\omega}(\bm{s},\bm{s}')$ peaks in the region  $|\bm{s}-\bm{s}'|\lesssim v_{F}/\omega\ll L$ and rapidly oscillates outside of it (see below). On the other hand, in such a  region, the local field  $E_{n}$ is nearly constant, i.e., $E_{n}(\bm{s})\approx E_{n}(\bm{s}')$, and so $Q_{s}$ takes the form
\begin{equation}
\label{Q-surface1}
Q_{s}=\frac{e^{2}\hbar^{4}}{4\pi m^{4} \omega^{3}} \! \int \!  dS  |E_{n}(\bm{s})|^{2} \bar{F}_{\omega}(\bm{s}),
\end{equation}
where $\bar{F}_{\omega}(\bm{s})=\! \int \!  dS'   F_{\omega}(\bm{s},\bm{s}')$ is, for $L\gg v_{F}/\omega$,  independent of the surface point $\bm{s}$.

Evaluation of $\bar{F}_{\omega}$ is based upon multiple-reflection expansion for the  electron Green function $G(\epsilon;\bm{s},\bm{s}')$ in a hard-wall cavity \cite{balian-ap70}. For   $L\gg \lambda_{F}$,  the direct and  single-reflection paths provide the dominant  contribution, while higher-order reflections are suppressed as powers of $\lambda_{F}/L\ll 1$ (see Appendix \ref{app:b}), and we obtain 
\begin{equation}
\label{rho}
\rho_{nn'}(\epsilon;\bm{s},\bm{s}')=2\nabla_{n}\nabla'_{n}\text{Im}G_{0}(\epsilon,\bm{s}-\bm{s}')
\end{equation}
%
where 
\begin{equation}
G_{0}(\epsilon,r)=\frac{m}{2\pi \hbar^{2}}\,\frac{e^{ik_{\epsilon}r}}{r},
~~~
k_{\epsilon}=\frac{\sqrt{2m\epsilon}}{\hbar},
\end{equation}
is  the free electron Green function, and factor 2 reflects  equal  contributions  of   direct and reflected paths at a surface point.  It is now  easy to see that, for $\epsilon\sim E_{F}$ and $\hbar\omega/E_{F}\ll 1$, the integrand of Eq.~(\ref{F}) peaks in the region  
\begin{equation}
|\bm{s}-\bm{s}'|\lesssim \, \frac{1}{k_{\epsilon+\hbar\omega}-k_{\epsilon}}\approx \frac{v_{F}}{\omega}
\end{equation}
and rapidly oscillates outside of it. This sets up the length scale $v_{F}/\omega$ for  correlation  function $F_{\omega}(\bm{s},\bm{s}')$ in Eq.~(\ref{Q-surface}) and leads  to Eq.~(\ref{Q-surface1}). The final step is to compute the normal derivatives in Eq.~(\ref{rho}) which, for $L\gg v_{F}/\omega$, is  accomplished by approximating the surface by the tangent plane at the surface point (see Appendix \ref{app:b}), yielding
\begin{equation}
\bar{F}_{\omega}=\hbar\omega \frac{2m^{4}E_{F}^{2}}{\pi \hbar^{8}}.
\end{equation}
Substituting this $\bar{F}_{\omega}$ into Eq.~(\ref{Q-surface1}), we finally arrive at  the surface contribution to the absorbed power
\begin{equation}
\label{Q-surface2}
Q_{s}
=\frac{e^{2}}{2\pi^{2} \hbar} \frac{E_{F}^{2}}{ (\hbar\omega)^{2}} \! \int \!  dS  |E_{n}|^{2}
=\frac{3v_{F}}{32\pi}\frac{\omega_{p}^{2}}{\omega^{2}}\! \int \!  dS  |E_{n}|^{2}.
\end{equation}
The above expression for the surface absorbed power $Q_{s}$, which is our central result, is valid for \textit{any} metal nanostructure with characteristic size $L\ll c/\omega$ in an alternating electric field with frequency $\omega\gg v_{F}/L$.

The surface contribution (\ref{Q-surface2})  should be considered in conjunction with the bulk contribution to the absorbed power. In fact, both contributions can be combined in the general expression (\ref{power-bulk}) for absorbed power by modifying the scattering rate in  the Drude dielectric function $\varepsilon(\omega)= \varepsilon_{i}(\omega)-\omega_{p}^{2}/\omega(\omega+i\gamma)$ as $\gamma=\gamma_{0}+\gamma_{s}$, where 
\begin{equation}
\label{rate-ld}
\gamma_{s}=
\frac{3 v_{F}}{4}
 \frac{\int \! dS |E_{n}|^{2}}{\int \! dV |\textbf{E}|^{2}},
\end{equation}
is the \textit{surface scattering rate}. Then, the surface contribution $Q_{s}$, Eq.~(\ref{Q-surface2}), is obtained  as the first-order term in the expansion of full absorbed power $Q$, Eq.~(\ref{power-bulk}),  over $\gamma_{s}$, implying that the surface scattering rate enters \textit{on par} with its bulk counterpart into the metal dielectric function. While $\gamma_{s}$ is  independent of the local field strength, it does depend strongly on its \textit{polarization} relative to the interface  and, in fact, represents the \textit{averaged}  over the surface  local scattering rate.

Finally, let us show that the  full plasmon decay rate $\Gamma$ due to \textit{both}  bulk and surface scattering is still given by the  general expression  (\ref{ld}), but  with  modified dielectric function $\varepsilon(\omega)$ that now includes the surface scattering rate  (\ref{rate-ld}). Indeed, using Eqs.~(\ref{Q-surface2}) and (\ref{energy-LL}), the surface contribution to $\Gamma$, i.e., the LD rate, takes the form 
\begin{equation}
\label{ld1}
\Gamma_{s} =  \frac{Q_{s}}{U}=  \frac{2\omega_{p}^{2}\gamma_{s}}{\omega^{3}}\left [\frac{\partial \varepsilon'(\omega)}{\partial \omega}\right ]^{-1} .
\end{equation}
The same expression is obtained by expanding Eq.~(\ref{ld}) [with modified $\varepsilon(\omega)$]  to the first order in $\gamma_{s}$.  For $\omega$ well below the interband transitions onset, the LD rate and surface scattering rate coincide, $\Gamma_{s}\approx\gamma_{s}$.

\section{Evaluation of surface scattering rates for specific geometries}
\label{sec:rates}

The strong polarization dependence of the surface absorbed power $Q_{s}$ and surface scattering rate $\gamma_{s}$ makes it possible to manipulate, in a wide range, the hot carrier excitation efficiency  by realigning the electric field orientation \cite{nordlander-nn15}.  This effect can be described by   \textit{surface enhancement factor} $M$  defined as the ratio of the full absorbed power, $Q=Q_{b}+Q_{s}$, to the bulk one, $Q_{b}$. Within RPA, the enhancement factor takes a simple form $M=1+\gamma_{s}/\gamma_{0}$, i.e., it is completely determined by the surface scattering rate.

Using our model,  surface scattering rates for  nanostructure of arbitrary shape can be evaluated directly from the local fields, without further resorting to quantum-mechanical calculations. In this section, we employ our main result Eq.~(\ref{rate-ld}) to evaluate  $\gamma_{s}$ for some common  structures: spherical particles, cylindrical wires, and spheroidal particles (nanorods and nanodisks). 

We start with recasting the surface scattering rate (\ref{rate-ld}) as the ratio of two surface integrals,
\begin{equation}
\label{rate-surface1}
\gamma_{s}=\frac{3  v_{F}}{4}\frac{\int \! dS |\nabla_{n}\Phi |^{2}}{\int \! dS \Phi^{*} \nabla_{n}\Phi},
\end{equation}
where the real part of the denominator is implied. This representation is especially useful for systems, whose shape permits separation of variables, and, as we show below, it yields \textit{analytical} results for some common structures, such as nanorods and nanodisks, which so far eluded any attempts of quantum-mechanical evaluation of $\gamma_{s}$.

\subsection{Spherical particles and cylindrical wires}

Let us first apply Eq.~(\ref{rate-surface1}) to the simplest case of a sphere of radius $a$. In the quasistatic limit, the potentials inside the sphere are given by regular solutions of the Laplace equation, $\Phi\propto  r^{l}Y_{lm}(\hat{\textbf{r}})$, where $r$ is the radial coordinate and $Y_{lm}(\hat{\textbf{r}})$ are the spherical harmonics ($l$ and $m$ are, respectively, the polar and azimuthal numbers). Then, a straightforward evaluation of Eq.~(\ref{rate-surface1}) recovers  the surface scattering rate for the $l$th mode \cite{lushnikov-zp74}:
\begin{equation}
\label{rate-sphere}
\gamma_{sp}^{l}=\frac{3l v_{F}}{4a}.
\end{equation}
The same rate is obtained for the $l$th transverse mode in an infinite cylindrical nanowire of radius $a$.

\subsection{Nanorods and nanodisks}

Nanorods and nanodiscs are often modeled by prolate and oblate spheroids, respectively. Here we distinguish between  longitudinal and transverse modes oscillating along the symmetry axis and within the symmetry plane, characterized, respectively, by   semiaxes $a$ and $b$  (see  Fig.~\ref{fig2}). Using  Eq.~(\ref{rate-surface1}), the surface scattering rate for all modes can be found in an analytical form (see Appendix \ref{app:c}), but here only the results for the dipole modes are presented.  For a  nanorod  (prolate spheroid) with aspect ratio $b/a<1$,  we obtain the following  rates for  longitudinal and transverse   polarizations, respectively:
\begin{align}
\label{rates-spheroid}
\gamma_{s}^{L} =\frac{3  v_{F}}{4a}\,
\frac{3}{2\tan^{2}\!\alpha}\left [\frac{2\alpha}{\sin 2\alpha}-1
\right ],
\nonumber\\
\gamma_{s}^{T} =\frac{3  v_{F}}{4a}\,
\frac{3}{4\sin^{2}\!\alpha}\left [1-\frac{2\alpha}{\tan 2\alpha}
\right ],
\end{align}
where $\alpha=\arccos (b/a)$ is the angular eccentricity. For a nanodisk (oblate spheroid) with  $b/a>1$, the rates (\ref{rates-spheroid})   apply with $\alpha=i\arccosh (b/a)$. Note that the CS rate for a spheroidal particle is \cite{schatz-jcp03} 
\begin{equation}
\gamma_{cs}=  \dfrac{v_{F} S}{4V}=\frac{3  v_{F}}{8a}\left (1 +  \frac{2\alpha}{\sin 2\alpha} \right ).
\end{equation}

\subsection{Numerical Results}

%
\begin{figure}[tb]
\begin{center}
\includegraphics[width=0.99\columnwidth]{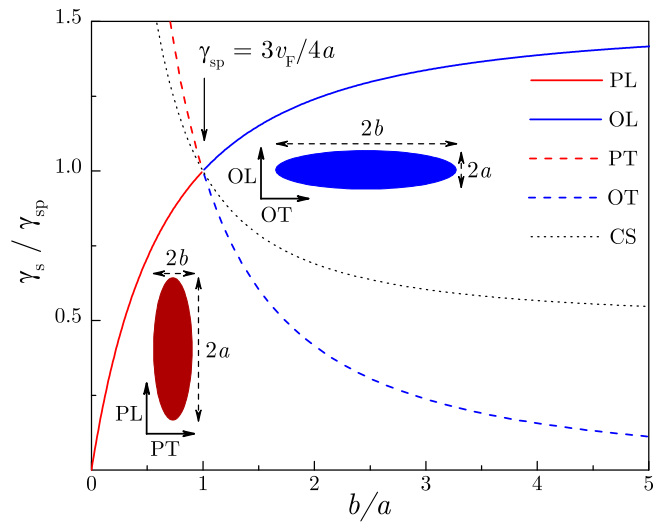}
\caption{\label{fig2}
Normalized surface scattering rates for prolate and oblate spheroids are  shown with changing aspect ratio $b/a$  along with the CS rate. Insets: Schematics of plasmon modes' polarizations.  
}
\end{center}
\vspace{-6mm}
\end{figure}
%

Here we present calculated surface scattering rates for spheroidal particles as the system shape evolves, with changing aspect ratio $b/a$, from a needle to a pancake. In Fig.~\ref{fig2}, we plot the rates (\ref{rates-spheroid})  normalized by the dipole mode rate $\gamma_{sp}=3v_{F}/4a$ for spherical particle of radius $a$.  At the sphere point $a=b$, the normalized rates for prolate and oblate spheroids  continuously transition into each other (e.g., PL to OL and PT to OT), and depending  on the mode polarization, exhibit dramatic differences in behavior with changing   aspect ratio. The normalized rate for the PL mode \textit{decreases} with reducing $b/a$, in sharp contrast to the CS rate, which shows the opposite trend. In the needle limit $b/a\ll 1$, the PL mode rate depends linearly on $b$,
\begin{equation}
\gamma_{s}^{\rm PL} \approx \frac{9\pi v_{F} b}{16a^{2}},
\end{equation}
while both the PT mode rate and CS rate are inversely proportional to $b$,
\begin{equation}
\gamma_{s}^{\rm PT} \approx \frac{9\pi v_{F}}{32b},
~~~~
\gamma_{cs} \approx \frac{3\pi v_{F}}{16b}.
\end{equation}
The similar behavior $\gamma_{s}^{\rm PT}$ and $\gamma_{cs}$ for $b/a\ll 1$ originates the fact that random ballistic scattering  is dominated by the shortest system length. Note, however, that the former exceeds the latter ($\gamma_{s}^{\rm PT}/\gamma_{cs}\rightarrow 3/2$) since directional scattering is more efficient than random one.

%
\begin{figure}[tb]
\begin{center}
\includegraphics[width=0.99\columnwidth]{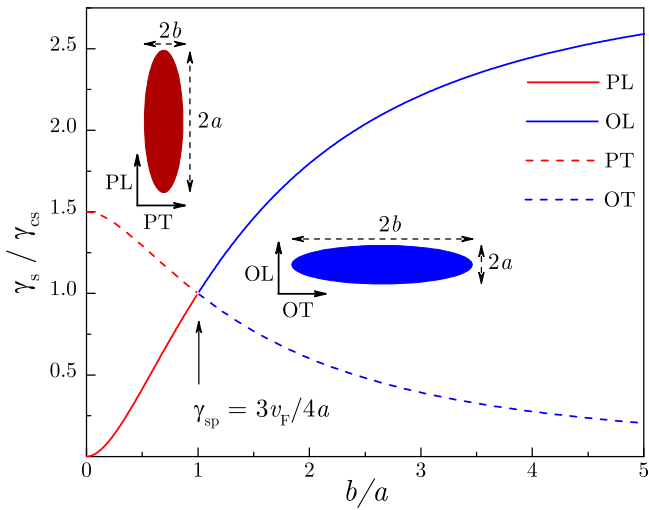}
\caption{\label{fig3}
Surface scattering rates for prolate and oblate spheroids normalized by the CS rate are  shown with changing aspect ratio $b/a$. 
}
\end{center}
\vspace{-6mm}
\end{figure}
%

For nanodisks ($b/a>1$), the above trends are reversed: With increasing $b/a$, as the nanodisk flattens, the normalized rates are increasing for the longitudinal (OL) mode and decreasing for the transverse (OT) mode (see Fig.~\ref{fig2}). In the pancake limit $b/a\gg 1$, the OL mode rate and CS rate are dominated by the pancake height $a$, which is now the shortest length,
\begin{equation}
\gamma_{s}^{\rm OL} \approx \frac{9 v_{F}}{8a},
~~~
\gamma_{cs}\approx \frac{3 v_{F}}{8a},
\end{equation}
with their ratio $\gamma_{s}^{\rm OL}/\gamma_{cs}\rightarrow 3$, while the OT mode rate exhibits a more complicated behavior: 
\begin{equation}
\gamma_{s}^{\rm OT} \approx \frac{ 9v_{F}a}{16b^{2}}\biggl [\ln\biggl(\frac{2b}{a}\biggr)^{2}-1\biggr ].
\end{equation}
To highlight the role of the local fields in surface scattering, we show in Fig.~\ref{fig3} the evolution, with changing $b/a$, of $\gamma_{s}$ for all modes, normalized by the CS  rate $\gamma_{cs}$. Here, we have $\gamma_{s}<\gamma_{cs}$ for the field polarization mostly tangential to the system boundary (PL and OT modes), and $\gamma_{s}>\gamma_{cs}$ for mostly normal polarization (PT and OL modes).  Note that recent measurements \cite{schatz-nl15} in cylinder-shaped nanorods and nanodisks revealed strong polarization dependence of the plasmon spectrum linewidth.

\section{Conclusions}
\label{sec:conc}

In conclusion, let us discuss the assumptions and approximations we made in deriving the surface scattering rate (\ref{rate-surface}). First, we assumed that the metal nanostructure is characterized by a single metal-dielectric interface. Our model can be straightforwardly extended to systems with two or more interfaces, such as, e.g., core-shell particles of various shapes or onion-like structures, by including each interface contribution in the matrix element $M_{\alpha\beta}^{s}$ [see Eq. (\ref{matrix})]. Importantly, the surface contribution $Q_{s}$ to the absorbed power, containing $|M_{\alpha\beta}^{s}|^{2}$ [see Eq.~(\ref{power-rpa})], will now include interference terms due to carrier scattering between the interfaces. For two interfaces that are sufficiently close to each other, such interference  terms would lead to \textit{coherent} oscillations (quantum beats), with the period $v_{F}/\omega$, of the surface scattering rate with changing interface separation. Such oscillations were recently identified and studied in detail for spherical metal nanoshells with dielectric core \cite{kirakosyan-prb16}.

We considered systems with characteristic size $L$  significantly larger than the nonlocality scale $v_{F}/\omega$ \cite{mortensen-pn13,mortensen-nc14} (i.e., with $L$ at least several nm large), and with electron level spacing much smaller than the optical energy $\hbar\omega$. Correspondingly, we disregarded quantum confinement effects that dominate optical response of small metal clusters. Specifically, the large  electron level spacing in nanometer-sized clusters leads to oscillatory behavior of the resonance width \cite{jalabert-prb02,jalabert-prb05} (not to be confused with the above coherent oscillations \cite{kirakosyan-prb16}) that should be visible, e.g., in small aspect ratio behavior of $\gamma_{s}$ in Figs. \ref{fig2} and \ref{fig3}. Such effects are best described within the TDLDA approach \cite{jalabert-prb02,jalabert-prb05,yuan-ss08,vallee-jpcl10,lerme-jpcc11,li-njp13,nordlander-acsnano14,mortensen-16}  and are out of  scope of this paper.

Finally, let us discuss  the effect of realistic confining potential profile on the surface scattering rate. While the hard-wall approximation is often used for systems larger than several nm \cite{kawabata-jpsj66,lushnikov-zp74,schatz-jcp83,barma-pcm89,yannouleas-ap92,eto-srl96,uskov-plasmonics13,khurgin-oe15}, recent  TDLDA calculations for spherical particles indicate that, even for relatively  large systems,  deviations of the surface barrier from rectangular shape do affect the overall magnitude of the plasmon decay rate  \cite{vallee-jpcl10,lerme-jpcc11,nordlander-acsnano14}. Importantly, the potential profile has distinct effects on the rapidly-oscillating electron wave-functions and slowly-varying plasmon local fields, which both determine the transition matrix element (\ref{matrix}). While, within TDLDA, the Kohn-Sham wave functions are directly determined by the (self-consistent) confining potential, the local fields are, instead, defined solely by the induced charge density via the (screened) Coulomb potential and, therefore,  depend on the confining potential  indirectly. Hence, the \textit{deviation} of  $E_{n}$ from its classical  behavior across the interface is determined by the electron density spillover over the classical (hard wall) boundary \cite{pustovit-prb06}, and, therefore,  is largely \textit{independent} of the system overall  shape.  Furthermore, recent TDLDA studies of relatively large (up to 10 nm) nanoparticles revealed \cite{vallee-jpcl10,lerme-jpcc11} that the main impact on plasmon linewidth comes precisely from the electron density tail and dielectric environment, implying that it is the plasmon local field near the interface, rather than electron wave functions, that chiefly determines the plasmon decay rate magnitude in real  structures. We now note that we employed the hard-wall approximation  \textit{only} for evaluation of the \textit{e-h} correlation function (\ref{F}), while retaining  explicit local field dependence in the surface scattering rate. Therefore, for general shape systems, the latter can still be obtained, in a good approximation,  from Eq.~(\ref{rate-surface}) using the classical local fields, but with the constant $A$ calculated self-consistently for some specific (e.g., spherical) system geometry, i.e., $A\approx 0.32$ \cite{vallee-jpcl10,lerme-jpcc11}.

In summary, we developed a quantum-mechanical theory for Landau damping of surface plasmons in metal nanostructures  of arbitrary shape. We derived an explicit expression for the surface scattering  rate that can be included, \textit{on par} with the bulk scattering rate, in the metal dielectric function. The rate is strongly dependent on the local field polarization, and is highly sensitive to the system geometry. Our results can be used for calculations of hot carrier generation rates in photovoltaics and photochemistry applications.


\acknowledgments
This work was supported in part by the National Science  Foundation under Grant  No. DMR-1610427 and No. HRD-1547754.

\appendix
\section{Transition matrix element}
\label{app:a}

To extract surface contribution to the matrix element 
\begin{equation}
M_{\alpha\beta} = \int\! dV \psi^{\ast}_{\alpha}\Phi \psi_{\beta},
\end{equation}
we first apply the Hamiltonian $H=-(\hbar^{2}/2m)\Delta$ to the wave function product as
\begin{align}
\label{id1}
 \psi^{\ast}_{\alpha}\psi_{\beta}
 &=
 \frac{1}{\epsilon_{\alpha \beta}} \left (\psi_{\beta} H\psi^{\ast}_{\alpha}-\psi^{\ast}_{\alpha} H \psi_{\beta}\right )
\nonumber\\
 &=
 \frac{\hbar^{2}}{2m\epsilon_{\alpha \beta}}\nabla_{\mu}  \left( \psi^{\ast}_{\alpha } \nabla_{\mu} \psi_{\beta} -  \psi_{\beta} \nabla_{\mu}\psi^{\ast}_{\alpha}\right),
\end{align}
where  $\epsilon_{\alpha\beta}=\epsilon_{\alpha}-\epsilon_{\beta}$ is the excitation energy, and summation over repeating indices $\mu=(x,y,z)$ is implied. After integrating by parts, the matrix element takes the form
\begin{align}
\label{matrixelement1}
M_{\alpha\beta} = \frac{e\hbar^{2}}{m\epsilon_{\alpha \beta}}\int \! dV  \psi^{\ast}_{\alpha} (\nabla_{\mu}\psi_{\beta}) E_{\mu},
\end{align}
where $eE_{\mu}=-\nabla_{\mu} \Phi$  are the electric field components, and we used that $\nabla_{\mu} E_{\mu}=0$ inside the metal and  $\psi_{\alpha}$ vanish at the boundary $S$.
Applying again the Hamiltonian $H$ to Eq. (\ref{matrixelement1}), we write
\begin{align}
\label{matrixelement2}
M_{\alpha\beta}
& = 
\frac{e\hbar^{2}}{m\epsilon_{\alpha \beta}^{2}}
\int \! dV \left [(H\psi^{\ast}_{\alpha}) \nabla_{\mu}\psi_{\beta}
-\psi^{\ast}_{\alpha}\nabla_{\mu} (H\psi_{\beta}) 
\right ] E_{\mu} 
\nonumber\\
&=
\frac{-e\hbar^{4}}{2m^{2}\epsilon_{\alpha \beta}^{2}}
\int \! dV \left [(\Delta\psi^{\ast}_{\alpha}) \nabla_{\mu}\psi_{\beta}
-\psi^{\ast}_{\alpha}\nabla_{\mu} (\Delta\psi_{\beta}) 
\right ] E_{\mu}.
\end{align}
Integrating the first term by parts yields the surface contribution
\begin{align}
\label{matrixelement3}
M_{\alpha\beta} ^{s}= 
\frac{-e\hbar^{4}}{2m^{2}\epsilon_{\alpha\beta}^{2}}
\int\!  dS_ {\nu}  (\nabla_{\nu}\psi^{\ast}_{\alpha})  E_{\mu} \nabla_{\mu}\psi_{\beta},
\end{align}
while the rest represents the bulk contribution, which can be manipulated to the form
\begin{align}
\label{matrixelement4}
M_{\alpha\beta}^{b}= 
\frac{e\hbar^{4}}{m^{2}\epsilon_{\alpha \beta}^{2}}
\int \! dV  (\nabla_{\mu}\psi^{\ast}_{\alpha}) (\nabla_{\nu} E_{\mu})  \nabla_{\nu}\psi_{\beta}.
\end{align}
Since the local fields change smoothly on the  Fermi wavelength scale, the bulk contribution is negligibly small. Noting that only \textit{normal} derivatives of $\psi_{\alpha}$  do not vanish at the (hard-wall) boundary  $S$, the surface contribution to  matrix element takes the form
\begin{equation}
\label{matrixelement5}
M_{\alpha\beta}^{s}= \frac{-e\hbar^{4}}{2m^{2} \epsilon_{\alpha\beta}^{2}} \! \int \! dS\,[\nabla_{n}\psi_{\alpha}(\bm{s})]^{*} E_{n}(\bm{s}) \nabla_{n}\psi_{\beta}(\bm{s}),
\end{equation}  
where $\nabla_{n} \psi_{\alpha}(\bm{s})\equiv\left [\bm{n}\cdot\bm{\nabla} \psi_{\alpha}(\bm{r})\right ]_{\bm{r}\rightarrow\bm{s_{-}}}$ is the wave function's normal derivative at the boundary point $\bm{s}$ on the inner side [$\bm{n}(\bm{s})$ is the outward normal to the surface at point $\bm{s}$] and $eE_{n}(\bm{s})=-\nabla_{n} \Phi(\bm{s}) $ defines the corresponding normal field component.


\section{Electron-hole surface correlation function}
\label{app:b}

\subsection{Multiple-reflection expansion}
 The electron Green function in a hard-wall potential well can be presented as an infinite series in reflections from the boundary as (suppressing energy dependence) \cite{balian-ap70}
\begin{equation}
G(\bm{r},\bm{r}')= G_{0}(\bm{r}-\bm{r}') - \frac{\hbar^{2}}{2m}\! \int\! dS G_{n}(\bm{r},\bm{s})G_{0}(\bm{s}-\bm{r}'),
\end{equation}
where $G_{n}(\bm{r},\bm{s})\equiv \nabla_{n} G(\bm{r},\bm{s})$ is the normal derivative of the Green function at surface point $\bm{s}$  on the boundary inner side, satisfying
\begin{equation}
\label{green-normal}
G_{n}(\bm{r},\bm{s})  =  2G_{n}^{0}(\bm{r}-\bm{s})-\frac{\hbar^{2}}{m}\! \int\!  \! dS'G_{n'}(\bm{r},\bm{s}')\bar{G}_{n}^{0}(\bm{s}'-\bm{s}).
\end{equation}
Here, $G_{0}(\epsilon,r)=(m/2\pi \hbar^{2})e^{ik_{\epsilon}r}/r$, with $k_{\epsilon}=\sqrt{2m\epsilon}/\hbar$, is the free electron Green function, and  
\begin{equation}
\bar{G}_{n}^{0}(\bm{s}'-\bm{s})=\frac{1}{2}\left [G_{n}^{0}(\bm{s}'-\bm{s}_{+}) +  G_{n}^{0}(\bm{s}'-\bm{s}_{-})\right ]
\end{equation}
is its \textit{symmetric} normal derivative at  the inner ($\bm{s}_{-}$) and outer ($\bm{s}_{+}$) boundary sides. Iterations of this system lead to the multiple-reflection expansion  \cite{balian-ap70}. For characteristic cavity size $L\gg \lambda_{F}$,  the leading contribution comes from the direct and single-reflection paths [first term in Eq. (\ref{green-normal})], while  the higher-order terms account for multiple reflections due to the surface curvature $R\sim L$, and are suppressed by powers of $\lambda_{F}/R$ \cite{balian-ap70}. Since the Fermi wavelength in metals is small, $\lambda_{F}< 1$ nm, the higher-order terms can be disregarded. The equation for $G_{nn'}(\bm{s},\bm{s}')$ is obtained by taking the normal derivative of Eq.~(\ref{green-normal}). Keeping only the first term, we obtain
\begin{equation}
G_{nn'}(\bm{s},\bm{s}')=2G_{nn'}^{0}(\bm{s}-\bm{s}').
\end{equation}

\subsection{Evaluation of $\bar{F}_{\omega}$}

To evaluate $\rho_{nn'}(\epsilon;\bm{s},\bm{s}')=2\text{Im}G_{nn'}^{0}(\bm{s}-\bm{s}')$, we use the fact that the size of characteristic region dominating surface integrals in the correlation function $F$ is  $|\bm{s}-\bm{s}'|\sim v_{F}/\omega\ll L$, and compute normal derivatives of $G_{0}(\bm{r}-\bm{r}')$ with respect to the tangent plane $z=0$,
\begin{align}
G_{zz'}(\epsilon,\bm{s}-\bm{s}')=2\left [\frac{\partial }{\partial z } \frac{\partial }{\partial z^{'}} G_{0}(\epsilon,\bm{r}-\bm{r}')\right ]_{z,z'=0}
\nonumber\\
=-2\left [\frac{\partial^{2}}{\partial z^{2}} G_{0}(\epsilon,\bm{r}-\bm{r}')\right ]_{z,z'=0}. 
\end{align}
Introducing notations $r=\sqrt{s^{2}+z^{2}}$, we write
\begin{equation}
\frac{\partial^{2}}{\partial z^{2}} G_{0}(\epsilon,r)=\left [\left (\frac{\partial r}{\partial z }\right )^{2}\frac{\partial^{2}}{\partial r^{2}} + \frac{\partial^{2} r}{\partial z^{2} }\frac{\partial }{\partial r} \right ] G_{0}(\epsilon,r),
\end{equation}
and, in the limit $z=0$, we obtain
\begin{equation}
\left [\frac{\partial^{2}}{\partial z^{2}}\, G_{0}(\epsilon,r)\right ]_{z=0} =\frac{1}{s} \frac{\partial }{\partial s} \, G_{0}(\epsilon,s),
\end{equation}
yielding
\begin{equation}
\rho_{zz}(\epsilon,s)=\frac{m}{\pi \hbar^{2} s} \frac{\partial }{\partial s} \, \frac{\sin k_{\epsilon}s}{s}.
\end{equation}
To evaluate $\bar{F}_{\omega}$, we note that for $L\gg v_{F}/\omega$, the  surface integral can be replaced by  integral over the tangent plane,
\begin{align}
\bar{F}_{\omega}
&=\frac{m^{2}}{\pi^{2} \hbar^{4}}\! \int\!  d\epsilon f_{\omega}(\epsilon)
\!  \int \!  \frac{d^{2}\bm{s}}{s^{2}}\left [\frac{\partial }{\partial s} \frac{\sin k_{\epsilon}s}{s}\right ]\! 
\left [\frac{\partial }{\partial s}  \frac{\sin k_{\epsilon+\hbar\omega}s}{s}\right ]
\nonumber\\
&=\frac{m^{2}}{4\pi \hbar^{4}}\! \int \! d\epsilon f_{\omega}(\epsilon)
\biggl [ k_{\epsilon} k_{\epsilon+\hbar\omega}  \left (k_{\epsilon}^{2}+k_{\epsilon+\hbar\omega}^{2} \right ) 
\\
&
~~~~~~~~~~~~~~~~~~~~~
 - \left (k_{\epsilon+\hbar\omega}^{2} -k_{\epsilon}^{2}\right )^{2} \arctanh \left (\frac{k_{\epsilon}}{k_{\epsilon+\hbar\omega}}\right )\biggr ].
 \nonumber
\end{align}
The function $f_{\omega}(\epsilon)=  f(\epsilon)-f(\epsilon+\hbar\omega)$ restricts the energy integral to the interval of width $\hbar\omega$, and, after rescaling the integration variable, we obtain
\begin{equation}
\bar{F}_{\omega}
=\hbar\omega \frac{2m^{4}E_{F}^{2}}{\pi \hbar^{8}}\,g(\hbar\omega/E_{F}),
\end{equation}
where the function
\begin{align}
g(\xi)=
&\int\limits_{-1/2}^{-1/2}\! dx \Biggl [(1+\xi x) \biggl [(1+\xi x)^{2}-\frac{\xi^{2}}{4}\biggr ]^{1/2}
\\
&
~~~~~~~~~~~
-\xi^{2}\arctanh\biggl [\frac{1+\xi(x-1/2)}{1+\xi(x+1/2)}\biggr ]^{1/2}
\Biggr ]
\nonumber
\end{align}
is normalized to unity, $g(0)=1$. Then, we obtain
\begin{equation}
\label{Q-surf2}
Q_{s}=\frac{e^{2}}{2\pi^{2} \hbar} \frac{E_{F}^{2}}{ (\hbar\omega)^{2}} \, g(\hbar\omega/E_{F}) \! \int \!  dS  |E_{n}|^{2}.
\end{equation}
Finally, for optical frequency well below the Fermi energy, $\hbar\omega/E_{F}\ll 1$, and using the relation $\omega_{p}^{2}=4\pi e^{2}n/m=4e^{2}k_{F}^{3}/3\pi m$, where $n$ is the electron concentration, we arrive at surface contribution to the absorbed power:
\begin{equation}
\label{Q-surf3}
Q_{s}=\frac{3  v_{F}}{32\pi}\frac{\omega_{p}^{2}}{\omega^{2}}\! \int \!  dS  |E_{n}|^{2}.
\end{equation}
%

\section{Scattering rate for separable shapes}
\label{app:c}

For system geometries that allow separation of variables, we present the potential  as $\Phi (\bm{r})=R (\xi)\Sigma(\eta,\zeta)$, where $\xi$ is the radial (normal) coordinate and the pair $(\eta,\zeta)$ parametrizes the surface. With surface area element $dS=h_{\eta}h_{\zeta} d\eta d\zeta$ and normal derivative $\nabla_{n}=h_{\xi}^{-1}(\partial/\partial\xi)$, where $h_{i}$ are the scale factors ($i=\xi,\eta,\zeta$), the surface scattering rate takes the form
\begin{equation}
\label{rate-sep}
\gamma_{s}=\frac{3  v_{F}}{4}\frac{R' (\xi)}{R (\xi)}
\frac{\int \! \int \! d\eta d\zeta (h_{\eta}h_{\zeta}/h_{\xi}^{2})  |\Sigma |^{2}}
{\int \! \int \! d\eta d\zeta (h_{\eta}h_{\zeta}/h_{\xi})  |\Sigma |^{2}}.
\end{equation}
Below we evaluate $\gamma_{s}$ for a spheroidal particle. 

Spheroidal  metal nanoparticles exhibit longitudinal and transverse plasmon modes with electric field oscillating, respectively, along the axis of symmetry  (semiaxis $a$) and within the symmetry plane (semiaxis $b$). Inside the prolate spheroid ($b/a<1$), the potential has the form $\Phi_{n}(\bm{r})\propto P_{l}^{|m|}(\xi)Y_{lm}(\eta,\phi)$, where $P_{l}^{m}(x)$ is the  Legendre function of the first kind. Spheroid surface corresponds to $\xi=a/f$ where $f=\sqrt{a^{2}-b^{2}}$ is half the distance between the foci, and the scale factors are given by 
\begin{align}
&h_{\xi}=f\sqrt{\frac{\xi^{2}-\eta^{2}}{\xi^{2}-1}},
~~ 
h_{\eta}=f\sqrt{\frac{\xi^{2}-\eta^{2}}{1-\eta^{2}}},
\nonumber\\
&h_{\phi}=f\sqrt{(\xi^{2}-1)(1-\eta^{2})}.
\end{align}
The surface area  and volume   of the prolate spheroid are
\begin{equation}
S=2\pi \left (b^{2}+\frac{ab\alpha}{\sin\alpha}\right ),
~~
V= \frac{4\pi}{3} b^{2}a,
\end{equation}
where $\alpha=\arccos (b/a)$ is the angular eccentricity. A straightforward evaluation of Eq. (\ref{rate-sep}) yields: 
\begin{align}
\gamma_{s}^{lm}=
&\frac{3  v_{F}}{4f}
\dfrac{(2l+1)!(l-|m|)!}{2(l+|m|)!}
\nonumber\\
&
~~~~~
\times
\frac{[P_{l}^{|m|}(\xi)]'}{P_{l}^{|m|}(\xi)}
\sqrt{\xi^{2}-1}
 \int  \limits_{-1}^{1} \! d\eta \dfrac{[P_{l}^{|m|}(\eta)]^{2}}{\sqrt{\xi^{2}-\eta^{2}}}.
\end{align}
%
%
For  longitudinal  and  transverse  \textit{dipole}  modes, i.e., $(lm)=(10)$ and $(lm)=(11)$, respectively, we obtain  $\gamma_{s}^{L,T}=(3  v_{F}/4a) f_{L,T}$, where
\begin{align}
f_{L}
=\frac{3}{2\tan^{2}\!\alpha}\left [\frac{2\alpha}{\sin 2\alpha}-1
\right ],
~
f_{T}
=\frac{3}{4\sin^{2}\!\alpha}\left [1-\frac{2\alpha}{\tan 2\alpha}
\right ],
\end{align}
are the  normalized (to spherical shape) rates. Within the CS model, the decay rate has the form $\gamma_{cs}= v_{F}S/4V=(3 v_{F}/4a)f_{cs}$, where
\begin{equation}
f_{cs}=\frac{aS}{3V}=\frac{1}{2}\left [1 +  \frac{2\alpha}{\sin 2\alpha} \right ].
\end{equation}
The rates for the oblate spheroid ($b/a>1$) are described by the above expressions  with  $\alpha=i\arccosh(b/a)$.



\begin{thebibliography}{99}

\bibitem{atwater-jap05} S. A. Maier and H. A. Atwater, 
J. Appl. Phys. \textbf{98}, 011101 (2005).

\bibitem{ozbay-science06} E. Ozbay, 
Science  \textbf{311}, 189 (2006).

\bibitem{stockman-review} M. I. Stockman, 
in \textit{Plasmonics: Theory and Applications}, edited by T. V. Shahbazyan and M. I. Stockman (Springer, New York, 2013).


%
\bibitem{sers} E. C. Le Ru and P. G. Etchegoin, \textit{Principles of Surface-Enhanced Raman Spectroscopy} (Elsevier, Amsterdam, 2009).
%


\bibitem{novotny-book} L. Novotny and B. Hecht, \textit{Principles of Nano-Optics} (CUP, New York, 2012).




\bibitem{bergman-prl03}
D. J. Bergman  and M. I. Stockman,
Phys. Rev. Lett.  \textbf{90}, 027402 (2003).





\bibitem{lakowicz-ab01} J. R. Lakowicz, 
Anal. Biochem. \textbf{ 298}, 1  (2001).



\bibitem{duyne-nm06}  J. Zhao, X. Zhang, C. Yonzon, A. J.   Haes, and R. P. Van Duyne, 
Nanomedicine \textbf{1}, 219 (2006).

\bibitem{atwater-nm10}H. A. Atwater and A. Polman,  
Nat. Mater. \textbf{9}, 205 (2010).



\bibitem{nordlander-nn15} M. L. Brongersma,	N. J. Halas, and P. Nordlander,
Nat. Nanotechnol. \textbf{10}, 25  (2015).



\bibitem{halperin-rmp86} W. P. Halperin, 
Rev. Mod. Phys. \textbf{58}, 533 (1986).

\bibitem{kresin-pr92} V. V. Kresin, 
Phys. Rep. \textbf{220},  1 (1992).


\bibitem{vallee-jpcb01}  C. Voisin, N. Del Fatti, D. Christofilos, and F. Vall\'{e}e,
J. Phys. Chem. B \textbf{105}, 2264  (2001).

\bibitem{schatz-jpcb03} K. L. Kelly, E. Coronado, L. L. Zhao, and G. C. Schatz, 
J. Phys. Chem. B \textbf{107} 668  (2003).


\bibitem{noguez-jpcc07} C. Noguez, 
J. Phys. Chem. C \textbf{111}  3806 (2007).


\bibitem{liao-prl82} A. Wokaun, J. P. Gordon, and P. F. Liao,
Phys. Rev. Lett. \textbf{48}, 957 (1982).





\bibitem{kreibig-book} U. Kreibig and M. Vollmer,
{\em Optical Properties of Metal Clusters}  (Springer, Berlin, 1995).




\bibitem{park-nl11} Y. K. Lee, C. H. Jung, J.Park, H. Seo, G. A. Somorjai, and J. Y. Park,
Nano Lett. \textbf{11}, 4251  (2011).

\bibitem{melosh-nl11} F. Wang and N. A. Melosh,  
Nano Lett. \textbf{11}, 5426  (2011).

\bibitem{halas-science11} M. W. Knight, H. Sobhani,  P. Nordlander,  and N. J. Halas,  
Science \textbf{332}, 702 (2011).

\bibitem{halas-nc13} A. Sobhani,	M. W. Knight,	Y. Wang,	B. Zheng,	N. S. King,	L. V. Brown, Z. Fang,	P. Nordlander, and N. J. Halas,
Nat. Commun. \textbf{4}, 1643 (2013).

\bibitem{lian-nl13} K. Wu,  W. E. Rodriguez-Cordoba, Y. Yang, and T. Lian, 
Nano Lett. \textbf{13}, 5255 (2013).

\bibitem{halas-nl13-2} M. W. Knight, Y. Wang, A. S. Urban, A. Sobhani, B. Y. Zheng, P. Nordlander, and N. J. Halas,
Nano Lett. \textbf{13}, 1687 (2013). 

\bibitem{clavero-np14} C. Clavero,
Nat. Photonics \textbf{8}, 95 (2014).



 
\bibitem{brongersma-nl14} H.  Chalabi, D. Schoen, M. L. Brongersma,  
Nano Lett. \textbf{14}, 1374  (2014).

\bibitem{atwater-nc14} R. Sundararaman, P. Narang, A. S. Jermyn, W. A. Goddard III, and H. A. Atwater,
Nat. Commun. \textbf{5}, 5788 (2014).

\bibitem{halas-nc15} B. Y. Zheng, H. Zhao, A. Manjavacas, M. McClain, P. Nordlander, and N. J. Halas,
Nat. Commun. \textbf{6}, 7797 (2015).



\bibitem{brongersma-nl11} I. Thomann, B. A. Pinaud, Z. Chen, B. M. Clemens, T. F. Jaramillo, and M. L. Brongersma,
Nano Lett.  \textbf{11}, 3440 (2011).

\bibitem{moskovits-nl12} J. Lee, S. Mubeen, X. Ji, G. D. Stucky, and M. Moskovits,
Nano Lett.  \textbf{12}, 5014 (2012).

\bibitem{halas-nl13} S. Mukherjee, F. Libisch, N. Large, O. Neumann, L. V. Brown, J. Cheng, J. B. Lassiter, E. A. Carter, P. Nordlander, and N. J. Halas,
Nano Lett. \textbf{13}, 240 (2013).

\bibitem{moskovits-nn13} S. Mubeen, J. Lee, N. Singh, S. Kr\"{a}mer, G. D. Stucky, and M. Moskovits,
Nat. Nanotechnol. \textbf{8}, 247 (2013).

\bibitem{halas-jacs14} S. Mukherjee, L. Zhou, A. M. Goodman, N. Large, C. Ayala-Orozco, Y. Zhang, P. Nordlander, and N. J. Halas,
J. Am. Chem. Soc. \textbf{136}, 64  (2014).

%




\bibitem{klar-prl98} T. Klar, M. Perner, S. Grosse, G. von Plessen, W. Spirkl, and J. Feldmann,
Phys. Rev. Lett. \textbf{80}, 4249 (1998).


\bibitem{mulvaney-prl02}C. S\"{o}nnichsen, T. Franzl, T. Wilk, G. von Plessen, J. Feldmann, O. V. Wilson, and P. Mulvaney,
Phys. Rev. Lett. \textbf{88}, 077402 (2002).

\bibitem{halas-prb02}S. L. Westcott, J. B. Jackson, C. Radloff, and N. J. Halas, 
Phys. Rev. B \textbf{66}, 155431 (2002).

\bibitem{klar-nl04}G. Raschke, S. Brogl, A. S. Susha, A. L. Rogach, T. A Klar, and J. Feldmann, 
Nano Lett. \textbf{4}, 1853 (2004).


\bibitem{vallee-prl04} A. Arbouet, D. Christofilos, N. Del Fatti, F. Vall\"{e}e, J. R. Huntzinger, L. Arnaud, P. Billaud, and M. Broyer, 
Phys. Rev. Lett. \textbf{93}, 127401 (2004).

\bibitem{halas-nl04}C. L. Nehl, N. K. Grady, G. P. Goodrich, F. Tam, N. J. Halas, and J. H. Hafner, 
Nano Lett. \textbf{4}, 2355 (2004).

\bibitem{hartland-pccp06} C. Novo, D. Gomez, J. Perez-Juste, Z. Zhang,   H.Petrova,   M. Reismann,   P. Mulvaney, and G. V. Hartland,  
Phys. Chem. Chem. Phys. \textbf{8}, 3540 (2006)


\bibitem{vallee-nl09} H. Baida, P. Billaud, S. Marhaba, D. Christofilos, E. Cottancin, A. Crut, J. Lerm\'{e}, P. Maioli, M. Pellarin, M. Broyer, N. Del Fatti, and F. Vall\'{e}e,
Nano Lett. \textbf{9}, 3463 (2009).



\bibitem{vanduyne-jpcc12} M. G. Blaber, A.-I. Henry, J. M. Bingham, G. C. Schatz, and R. P. Van Duyne,
J. Phys. Chem. C \textbf{116}, 393 (2012).

\bibitem{vallee-nl13} V. Juv\'{e}, M. F. Cardinal, A. Lombardi, A. Crut, P. Maioli, J. P\"{e}rez-Juste, L. M. Liz-Marz\'{a}n, N. Del Fatti, and F. Vall\'{e}e,
Nano Lett.  \textbf{13}, 2234 (2013).

\bibitem{schatz-nl15} M. N. O'Brien, M. R. Jones, K. L. Kohlstedt, G. C. Schatz, and C. A.
Mirkin, 
Nano Lett. \textbf{15}, 1012 (2015).


\bibitem{kawabata-jpsj66} A. Kawabata and R. Kubo,  
J. Phys. Soc. Jpn. \textbf{21}, 1765 (1966).

\bibitem{lushnikov-zp74} A. A. Lushnikov and A. J. Simonov, 
Z. Physik \textbf{270}, 17 (1974).


\bibitem{schatz-jcp83} W. A. Kraus and G. C. Schatz,
J. Chem. Phys. \textbf{79}, 6130 (1983).

\bibitem{barma-pcm89} M. Barma and V. J. Subrahmanyam,
J. Phys.: Cond. Mat. \textbf{1}, 7681 (1989).

\bibitem{yannouleas-ap92} C. Yannouleas and R. A. Broglia,  
Ann. Phys. \textbf{217}, 105 (1992).

\bibitem{eto-srl96} M. Eto and K. Kawamura,
Surf. Rev. Lett. \textbf{3}, 151 (1996).

\bibitem{uskov-plasmonics13} A. V. Uskov,  I. E. Protsenko, N. A. Mortensen, and E. P. O'Reilly,
Plasmonics \textbf{9}, 185 (2013).

\bibitem{khurgin-oe15} J. B. Khurgin and G. Sun,
Opt. Express \textbf{23}, 250905 (2015).



\bibitem{jalabert-prb02} R. A. Molina, D. Weinmann, and R. A. Jalabert,
Phys. Rev. B \textbf{65}, 155427 (2002).

\bibitem{jalabert-prb05} G. Weick, R. A. Molina, D. Weinmann, and R. A. Jalabert,
Phys. Rev. B \textbf{72}, 115410 (2005).

\bibitem{yuan-ss08} Z. Yuan and S. Gao,
Surf. Sci.  \textbf{602}, 460 (2008).

\bibitem{vallee-jpcl10} J. Lerm\'{e}, H. Baida, C. Bonnet, M. Broyer, E. Cottancin, A. Crut, P. Maioli, N. Del Fatti, F. Vall\'{e}e, and M. Pellarin,
J. Phys. Chem. Lett. \textbf{1}, 2922 (2010).

\bibitem{lerme-jpcc11} J. Lerm\'{e},
J. Phys. Chem. C \textbf{115}, 14098 (2011).


\bibitem{li-njp13} X. Li, Di Xiao, and Z. Zhang,
New J. Phys. \textbf{15}, 023011 (2013).


\bibitem{nordlander-acsnano14} A. Manjavacas, J. G. Liu, V. Kulkarni, and P. Nordlander,
ACS Nano \textbf{8}, 7630 (2014).

\bibitem{mortensen-16} 
T. Christensen, W. Yan, A.-P. Jauho, M. Solja\v{c}i\'{c}, and N. A. Mortensen,
arXiv:1608.05421.



\bibitem{mortensen-pn13} N. A. Mortensen, 
Photonic. Nanostruct. \textbf{11}, 303 (2013).

\bibitem{mortensen-nc14} N. A. Mortensen,	S. Raza,	M. Wubs,	T. S{\o}ndergaard, and S. I. Bozhevolnyi, 
Nat. Commun. \textbf{5}, 3809 (2014).



\bibitem{kreibig-zp75} L. Genzel, T. P. Martin, and U. Kreibig, 
Z. Phys. B \textbf{21}, 339 (1975).

\bibitem{ruppin-pss76} R. Ruppin and H. Yatom,
Phys. Status Solidi \textbf{74}, 647 (1976).

\bibitem{schatz-cpl83} W. A. Krauss and G. C. Schatz, 
Chem. Phys. Lett. \textbf{99}, 353 (1983).

\bibitem{schatz-jcp03}
E. A. Coronado and G. C. Schatz,
 J. Chem. Phys. \textbf{119}, 3926 (2003).

\bibitem{moroz-jpcc08} A. Moroz, 
J. Phys. Chem. C \textbf{112}, 10641 (2008).


\bibitem{dionne-nature12} J. A. Scholl, A. L. Koh,  J. A. Dionne,  
Nature (London) \textbf{483}, 421 (2012) .



\bibitem{kirakosyan-prb16} A. S. Kirakosyan, M. I. Stockman, and T. V. Shahbazyan,  
Phys. Rev. B \textbf{94}, 155429 (2016). 



\bibitem{shahbazyan-prl16} T. V. Shahbazyan, 
Phys. Rev. Lett. \textbf{117}, 207401 (2016).

\bibitem{landau}  L. D. Landau and E. M. Lifshitz, {\it Electrodynamics of Continuous Media} (Elsevier, Amsterdam, 2004). 

\bibitem{mahan-book} G. D. Mahan, {\it Many-Particle Physics} (Plenum, New York, 1990).

\bibitem{balian-ap70} R. Balian and C. Bloch, 
Ann. Phys. \textbf{60}, 401 (1970).

\bibitem{pustovit-prb06} V. N. Pustovit and T. V. Shahbazyan,
Phys. Rev. \textbf{B} 73, 085408 (2006)

\end{thebibliography}
\end{document}